\documentclass{article}

\usepackage{arxiv}

\usepackage[utf8]{inputenc} % allow utf-8 input
\usepackage[T1]{fontenc}    % use 8-bit T1 fonts
\usepackage{hyperref}       % hyperlinks
\usepackage{url}            % simple URL typesetting
\usepackage{booktabs}       % professional-quality tables
\usepackage{amsfonts}       % blackboard math symbols
\usepackage{nicefrac}       % compact symbols for 1/2, etc.
\usepackage{microtype}      % microtypography
\usepackage{lipsum}
\usepackage{graphicx}
\usepackage{glossaries}
\usepackage{caption}
\usepackage{subcaption}
\usepackage{float}
\usepackage{cleveref}
\usepackage{array}   % Paket für bessere Tabellengestaltung
\usepackage{amssymb}
\usepackage{textcomp}
\usepackage{siunitx}

\usepackage{arxiv}
\usepackage{authblk} % for \affil command
\graphicspath{ {./images/} }

\crefname{figure}{Fig.}{Figs.}
\crefname{section}{Sec.}{Secs.}
\crefname{table}{Tab.}{Tabs.}

\title{EAP4EMSIG - Enhancing Event-Driven Microscopy for Microfluidic Single-Cell Analysis}

\author[1,2,*]{Nils~Friederich}
\author[1,*]{Angelo~Jovin~Yamachui~Sitcheu}
\author[1,2]{Annika~Nassal}
\author[3]{Erenus~Yildiz}
\author[4]{Matthias~Pesch}
\author[1]{Maximilian~Beichter}
\author[3]{Lukas~Scholtes}
\author[1]{Bahar~Akbaba}
\author[1]{Thomas~Lautenschlager}
\author[1]{Oliver~Neumann}
\author[4]{Dietrich~Kohlheyer}
\author[3]{Hanno~Scharr}
\author[4,5,$\oplus$]{Johannes~Seiffarth}
\author[4,$\oplus$]{Katharina~Nöh}
\author[1,$\oplus$]{Ralf~Mikut}

\affil[1]{Institute for Automation and Applied Informatics (IAI), Karlsruhe Institute of Technology}
\affil[2]{Institute of Biological and Chemical Systems (IBCS), Karlsruhe Institute of Technology}
\affil[3]{Institute for Data Science and Machine Learning (IAS-8), Forschungszentrum Jülich GmbH}
\affil[4]{Institute of Bio- and Geosciences (IBG-1), Forschungszentrum Jülich GmbH}
\affil[5]{Computational Systems Biology (AVT-CSB), RWTH Aachen University}
\affil[*]{Contributed Equally}
\affil[$\oplus$]{Supervised Equally}

\newcommand{\cmmnt}[1]{}
\newcolumntype{Y}{>{\centering\arraybackslash}X}

\makeglossaries
\newacronym{ai}{AI}{Artificial~Intelligence}

\newacronym[%
  shortplural={APs},%
  longplural={Average~Precisions}%
] {ap}{AP}{Average~Precision}

\newacronym[
  shortplural={CNNs},%
  longplural={Convolutional~Neural~Networks}%
]{cnn}{CNN}{Convolutional Neural Network}

\newacronym{cpn}{CPN}{Contour~Proposal~Network}

\newacronym{cpu}{CPU}{Central Processing Unit}

\newacronym{db}{DB}{DataBase}

\newacronym{dl}{DL}{Deep~Learning}

\newacronym{dora}{DORA}{Dataflow-Oriented~Robotic~Architecture}

\newacronym{eap}{EAP}{Experiment~Automation~Pipeline}

\newacronym{eap4emsig}{EAP4EMSIG}{Experiment~Automation~Pipeline~for~Event-Driven~Microscopy~to~Smart~Microfluidic~Single-Cell~Analysis}

\newacronym{eapdp}{EAPDP}{Experiment~Automation~Pipeline~for~Dynamic~Processes}

\newacronym[%
  shortplural={FMs},%
  longplural={Foundation Models}%
]{fm}{FM}{Foundation Model}

\newacronym[
    shortplural={GPUs},%
    longplural={Graphics Processing Units}%
]{gpu}{GPU}{Graphics Processing Unit}

\newacronym{gui}{GUI}{Graphical~User~Interface}

\newacronym{kaida}{KaIDA}{Karlsruhe~Image~Data~Annotation}

\newacronym{log}{LoG}{Laplacian of Gaussian}

\newacronym{mae}{MAE}{Mean~Absolute~Error}

\newacronym[%
  shortplural={mAPs},%
  longplural={mean~Average~Precisions}%
] {map}{mAP}{mean~Average~Precision}

\newacronym{ml}{ML}{Machine~Learning}

\newacronym{mlci}{MLCI}{Microfluidic~Live-Cell~Imaging}

\newacronym{mlp}{MLP}{Multi-Layer~Perceptron}

\newacronym{mp}{MP}{Mixed~Precision}

\newacronym{nn}{NN}{Neural~Network}

\newacronym{omero}{OMERO}{Open~Microscopy~Environment~Remote~Objects}

\newacronym{poc}{PoC}{Proof~of~Concept}

\newacronym{pq}{PQ}{Panoptic~Quality}

\newacronym{pyme}{PYME}{PYthon~Microscopy~Environment}

\newacronym{ram}{RAM}{Random-Access~Memory}

\newacronym{relu}{ReLU}{Rectified~Linear~Unit}

\newacronym{ros}{ROS}{Robot~Operating~System}

\newacronym{rq}{RQ}{Recognition~Quality}

\newacronym{sam}{SAM}{Segment~Anything~Model}

\newacronym{sota}{SOTA}{State-Of-The-Art}

\newacronym{sq}{SQ}{Segmentation~Quality}

\newacronym{tp}{TP}{True~Positive}

\newacronym{ui}{UI}{User~Interface}

\newacronym[%
  shortplural={VFMs},%
  longplural={Vision Foundation Models}%
]{vfm}{VFM}{Vision Foundation Model}

\newacronym{yolo}{YOLO}{You~Only~Look~Once}

\begin{document}
\maketitle
\begin{abstract}
\gls{mlci} yields data on microbial cell factories.  However, continuous acquisition is challenging as high-throughput experiments often lack real-time insights, delaying responses to stochastic events. We introduce three components in the \gls{eap4emsig}: a fast, accurate \gls{mlp}-based autofocusing method predicting the focus offset, an evaluation of real-time segmentation methods and a real-time data analysis dashboard. Our \gls{mlp}-based autofocusing achieves a \gls{mae} of \SI{0.105}{\micro\meter} with inference times from \SI{87}{\milli\second}. Among eleven evaluated \gls{dl} segmentation methods, Cellpose reached a \gls{pq} of \SI{93.36}{\percent}, while a distance-based method was fastest (\SI{121}{\milli\second}, Panoptic Quality \SI{93.02}{\percent}).
\end{abstract}

\section{Introduction}
\label{section:introduction}
Microorganisms are ubiquitous and diverse life forms that play a crucial role in ecological processes, human health, and industrial applications~\cite{AmericanSocietyforMicrobiology.2022}. The study of their behavior, genetics and interactions at the single-cell level is therefore of fundamental importance for advancements in biotechnology, environmental protection, and medicine~\cite{Nduko.2020, Ganesan.2022}. \gls{mlci} has established itself as a powerful technology to observe dynamic processes in microbial populations with high spatial and temporal resolution, thereby generating large amounts of experimental data~\cite{r2024microfluidic}. However, the effective use of this technology in high-throughput experimental setups, where thousands of microcultures are usually analyzed in parallel, poses technical challenges~\cite{friederich2024eap4emsig}. Therefore, real-time responsiveness is crucial because delayed reactions to biological events (e.g., sudden changes in cell growth, cell stress responses) or technical issues (e.g., focus drift, fluidic disturbances) can irreversibly compromise data quality or even result in the failure of the entire experiment. 

To perform such a \gls{mlci}, microfluidic chips are used, which contain thousands of individual chambers with parallel-grown microbe colonies. Microbes are typically injected into these chambers at the beginning of the experiment. Due to variability in production, each chip can have a slightly different shape in z, as well as in x and y. While variations in the x- and y-dimensions of the chip can be compensated by detecting the chamber via real-time \gls{dl} methods, inaccuracies along the z-axis require a dedicated focusing method. In terms of our high-throughput experiment, a real-time one is essential because a loss of focus could make it partially or entirely infeasible to recognize the microorganisms. Therefore, autofocusing per chamber is essential for maximum information extraction. However, existing autofocusing techniques either rely on specialized hardware, which limits their flexibility and increases costs, or depend on computationally expensive image-based methods that are unsuitable for real-time applications.

After refocusing, normally, the chamber would be extracted via the mentioned detection method and the microbes could then be extracted each via \gls{sota} Segmentation methods. But current \gls{sota} methods, while highly accurate, often require extensive computational resources, making them impractical for rapid decision-making in high-throughput microfluidic experimental setups. 

If the microbes are extracted, they can be analyzed in time-independent features like position, cell size, state (lively/dead) and time-dependent features like growth rate. This information must be made available to the biologist conducting the experiment so that they can generate live findings and adjust the experiment settings if necessary. A significant limitation is the lack of integrated systems that enable comprehensive real-time analysis of the acquired data and allow researchers to react directly to relevant biological or technical "events". Such events can include critical phases in the cell cycle, cellular stress responses, morphological changes, as well as technical malfunctions such as focus loss or problems with fluidics~\cite{shroff2024live}. This lack of direct feedback and control capabilities delays the adaptive optimization of experiments and the acquisition of more profound insights, especially when investigating stochastic cellular processes.

This vision of Smart Microfluidic Single-Cell Analysis, therefore, requires event-driven microscopy, where the system autonomously or semi-autonomously reacts to detected events. This represents an advancement over the established workflow, where data is often analyzed only post-experimentally. To realize this vision, a robust automation pipeline is needed that integrates fast and intelligent components for image acquisition, processing and analysis. Current solutions in experiment automation~\cite{chiron2022cybersco,fox2022enabling,mahecic2022event,pedone2021cheetah,pinkard2021pycro,susano2021python,pyme, friederich2023ai} often cover only partial aspects or are not sufficiently modular and adaptive for the specific requirements of \gls{mlci} with a focus on event-driven approaches.

To achieve this, we are building on our work \gls{eap4emsig}~\cite{friederich2024eap4emsig} and improving the autofocus, segmentation, and real-time dashboard modules mentioned above:

\begin{enumerate}
    \item A novel, \gls{mlp}-based real-time autofocusing method designed to ensure continuously high image quality: In contrast to existing approaches, our proposed autofocusing method leverages a simplified \gls{mlp} architecture that directly predicts the optimal focus offset from rapidly computable image features, thereby achieving both high-precision and real-time performance without the need for additional hardware.

    \item A comprehensive evaluation of \gls{sota} \gls{dl} segmentation methods to identify suitable models for event-driven analysis: Since developing a new segmentation method is beyond the scope of this paper, we instead perform a detailed comparative evaluation of existing \gls{sota} segmentation models—task-specific, domain-specific, and foundation models—to identify the most suitable approach balancing accuracy and computational cost for real-time applications in~\gls{mlci}.

    \item A new real-time data analysis dashboard that allows biologists to monitor ongoing experiments in detail, visualize analysis results in real-time, and offers enhanced control options to react directly to detected events. All proposed methods and evaluations are rigorously validated through extensive benchmarking on representative microbial image datasets, ensuring the robustness and practical relevance of the results.
\end{enumerate}

By integrating these components, we aim to enhance automated, event-driven analysis of microbial single cells. In the following, we will describe the methodology of these three components in detail, evaluate their performance, and discuss their importance in the context of intelligent experiment automation.

\section{Related Work}
\label{section:related_work}

The successful automation of \gls{mlci} for insightful, high-throughput single-cell analysis of microbes hinges on overcoming challenges in experiment control, image acquisition, and real-time data processing. This section reviews the \gls{sota} in these interconnected domains, starting with \glspl{eap} and the emergence of event-driven microscopy. We then delve into specific challenges and advances in real-time autofocusing and real-time image segmentation, critically evaluating existing approaches to highlight the research gaps that motivate the contributions of this paper.

\subsection{Experiment Automation Pipeline and Rise of Event-Driven Microscopy}
\label{subsection:related_work-eap_tools}

\begin{table*}[ht]
    \centering
    \caption{Overview of \gls{sota} \gls{eap} methods based on the implemented modules along with their limitations regarding the context of this work. \acrfull{pyme}, \acrfull{eapdp}.}
    \begin{tabular}{c|c|c|c|c|c}
        \toprule
        Method & \multicolumn{4}{|c|}{Modules} & Limitations\\
        \midrule
            & Microscope & Real-time & Real-time & Real-time \\
            & control& image & data & experiment \\
            & & processing & analysis & planner\\
        \midrule
        CyberSco.Py~\cite{chiron2022cybersco} & \checkmark  & \checkmark & \checkmark & \checkmark  & Supports only U-Net \\
         MicroMator~\cite{fox2022enabling} & \checkmark  & \checkmark & \checkmark & \checkmark & Not actively used\\
         Event-driven acquisition~\cite{mahecic2022event} & \checkmark  & \checkmark  &  & \checkmark & Other modules absent\\
        Cheetah~\cite{pedone2021cheetah} & \checkmark  & \checkmark & \checkmark & \checkmark  & Supports only U-Net \\
        Pycro-Manager~\cite{pinkard2021pycro} & \checkmark  & \checkmark & \checkmark &   & Other modules absent \\
        Python-Microscope~\cite{susano2021python} & \checkmark  & & \checkmark &  &  Other modules absent \\
        \acrshort{pyme}~\cite{pyme} & \checkmark  & \checkmark & \checkmark &   &  Focus on super-resolution\\
        \acrshort{eapdp}~\cite{friederich2023ai} & \checkmark  & \checkmark & \checkmark & \checkmark &  Focus on dynamic process modelling\\
        \bottomrule
    \end{tabular}
    \label{tab:eap_sota}
\end{table*}
\glspl{eap} are becoming increasingly vital in life science research to enhance reproducibility, increase throughput, and manage complex experimental workflows~\cite{holland2020automation}. In the context of \gls{mlci}, which generates vast quantities of image data, robust automation is essential~\cite{10.1063/5.0170050, huang2024transformative}.

Recently, the paradigm of event-driven microscopy has gained traction, aiming to make experimental systems more intelligent and responsive. Several efforts have sought to incorporate such principles. For instance, tools like CyberSco.Py~\cite{chiron2022cybersco} and MicroMator~\cite{fox2022enabling} have pioneered event-based conditional microscopy and reactive microscopy workflows, respectively, allowing the system to adapt acquisition based on detected image features. Dedicated systems for event-driven acquisition ~\cite{mahecic2022event} have further demonstrated the potential to enrich experimental datasets by intelligently responding to observed phenomena.

However, while these event-driven systems showcase the value of responsive experiment control, they often focus on specific aspects of the image acquisition workflow or possess certain limitations when considering a holistic, innovative single-cell analysis pipeline for \gls{mlci}. For example, some systems might be tailored to particular hardware setups, offer limited modularity for easily integrating diverse, new analytical tools (such as the here presented advanced real-time autofocusing and versatile segmentation capabilities required for immediate downstream decision-making), or might not provide comprehensive, user-friendly dashboards designed for biologist-in-the-loop interaction during complex experiments. A significant hurdle often remains in the seamless integration of rapid image processing (for both event detection and quantitative analysis) with immediate feedback to microscope control and interactive data visualization.

Beyond these specialized event-driven systems, a broader range of \gls{eap} tools relevant to microscopy exists (see Table 1 for an overview, referencing~\cite{friederich2023ai} and its content). While tools for cybergenetic control (e.g., Cheetah~\cite{pedone2021cheetah}) advance control paradigms, and platforms like \gls{pyme}~\cite{pyme} support specialized niches such as super-resolution, many general microscope control software packages (e.g., Pycro-Manager~\cite{pinkard2021pycro}, Python-Microscope~\cite{susano2021python}) provide essential low-level control but often lack the integrated high-level intelligence for event-driven experiment planning or sophisticated real-time data analysis specifically tailored to single-cell \gls{mlci}. Even more comprehensive pipeline approaches like \gls{eapdp}~\cite{friederich2023ai} have primary objectives differing from the specific, combined needs of smart, event-driven microbial single-cell analysis, which requires rapid, adaptive image acquisition coupled with immediate on-the-fly processing, event detection and visualization.

To the best of our knowledge, despite these valuable contributions, there is still a clear lack of a complete, modular, extendable, and adaptable pipeline in the field of \gls{mlci} that specifically and seamlessly integrates rapid, intelligent, and hardware-agnostic autofocus; robust, real-time capable segmentation suitable for reliable event triggering and quantitative analysis; and an intuitive, interactive dashboard for real-time monitoring, in-depth analysis and event-based intervention capabilities~\cite{friederich2024eap4emsig}. This identified gap strongly motivates the continued development and enhancement of our \gls{eap4emsig} system and the specific contributions presented in this paper.

\subsection{Challenges and Advances in Automated Autofocusing for Event-Driven MLCI}
\label{subsection:related_work-autofocusing}
The successful execution of event-driven \gls{mlci}, as motivated above, is critically dependent on consistently acquiring high-quality images. Maintaining optimal focus over many positions and long time periods is essential, especially when imaging microbial cells in dynamic microfluidic chambers. These setups can be challenging due to low contrast, changing cell densities, and varying media conditions. Traditional autofocusing approaches are broadly categorized in hardware-based and image-based~\cite{ li2002autofocus,bian2020autofocusing}.

Hardware-based methods, utilizing additional sensors~\cite{li2002autofocus}, can offer precision and speed. However, they typically increase system complexity and cost~\cite{bian2020autofocusing} and may lack the flexibility to easily integrate into highly adaptable, software-centric event-driven workflows.

Image-based methods analyze image content to find the optimal focal plane using various metrics~\cite{sun2004autofocusing}. While cost-effective, they often struggle with the low-contrast samples typical of some microbial cultures and can be computationally intensive, especially if requiring the processing of entire Z-stacks for each focusing event~\cite{shih2007autofocus}. Furthermore, a critical limitation of traditional metrics, such as those based on image gradients (e.g., Sobel, Laplacian variance) or contrast, is their inability to determine the direction of the required focus adjustment. These methods can measure image sharpness but cannot distinguish between a positive and a negative focus deviation. This lack of directionality makes them unsuitable for rapid, closed-loop convergence, as the system would oscillate without knowing whether to move the objective up or down. For these reasons, a learning-based paradigm that can predict both the magnitude and direction of the focus offset is essential for robust, real-time control. This computational load frequently renders them too slow (i.e., exceeding latency requirements of tens of milliseconds) for the rapid, iterative focus adjustments essential in real-time, event-driven \gls{mlci}.

\gls{ml}, particularly \gls{dl}, offers promising alternatives by enabling models to predict optimal focus directly from image data, potentially from single frames~\cite{herrmann2020learning,liao2021deep}. While \gls{nn} models can be integrated into automation pipelines~\cite{huang2024transformative}, challenges include the need for extensive annotated datasets for training complex architectures (e.g., deep \glspl{cnn} or Transformers) and ensuring robust performance on novel organisms or experimental variations~\cite{padovani2022segmentation}. Furthermore, many existing \gls{ml}-based autofocus methods might still rely on Z-stack analysis or employ models whose complexity does not meet the stringent low-latency requirements of an event-driven system. This highlights the need for lightweight, extremely fast (e.g., <\SI{100}{\milli\second} prediction time), yet accurate \gls{ml} models for focus offset prediction, like the \gls{mlp}-based approach proposed in this paper, which can operate efficiently without specialized hardware, ideally using features from rapidly acquired images to predict the focus offset directly.

\subsection{Real-time Image Segmentation for Microbial Single-Cell Analysis in Event-Driven Workflows}
\label{subsection:related_work-segmentation}

For event-driven \gls{mlci} to enable smart single-cell analysis, the rapid and accurate segmentation of individual microbial cells within potentially dense and structurally complex colonies is fundamental. This enables the extraction of quantitative single-cell data (e.g., cell size, morphology, growth rate) and the real-time detection of cellular events, which can then trigger experimental interventions.
Classic image processing methods~\cite{lewis1995fast, otsu1975threshold} are generally ill-suited for these tasks because they require manual feature engineering and parameter tuning. They also struggle with changes in imaging conditions and cell appearance~\cite{wang2022medical}, making them impractical for high-throughput, real-time applications. 
\gls{dl}-based methods have become the \gls{sota} for most biomedical image segmentation tasks due to their superior accuracy and robustness~\cite{7881449}. For preliminary tasks such as detecting the region of interest (e.g., the growth chamber itself), established object detection models like \gls{yolo}~\cite{redmon2016you} offer strong performance and speed~\cite{friederich2024security} and can be readily integrated. However, for the more challenging task of segmenting individual microbial cells, a careful selection of models is required:

\begin{itemize}
    \item \textbf{Task-specific models:}
    Approaches like the distance-based method~\cite{scherr2020cell} trained on data from microbes similar to ours~\cite{scherr2022microbeseg} (see \cref{section:experiments}) and Omnipose~\cite{cutler2022omnipose} are often highly optimized for cell segmentation. While achieving excellent accuracy, their performance in terms of raw speed for true real-time feedback (i.e., sub-100ms processing), generalization to microbial species significantly different from their training data, and ease of deployment in integrated event-driven systems need to be critically assessed for each specific \gls{mlci} application.

    \item \textbf{Domain-specific models:}
    Well-known models like StarDist~\cite{schmidt2018cell}, Cellpose 3~\cite{Stringer2024.02.10.579780}, and the \gls{cpn}~\cite{celldetection} offer good generalization in biomedical imaging. However, their balance of accuracy versus computational cost is a key concern for real-time microbial \gls{mlci}. Their out-of-the-box performance on specific challenges, such as segmenting low-contrast microbes or distinguishing individual cells in extremely dense colonies under rapid imaging conditions, may vary and thus warrants investigation.

    \item \textbf{Foundation Models (FMs)}
    Large-scale models like \gls{sam}, incl. its variants~\cite{kirillov2023segany,ravi2024sam}, Florence-2~\cite{xiao2024florence}, BiomedParse~\cite{zhao2024biomedparse}, and 4M21~\cite{4m21}, offer impressive generalization. However, for precise, real-time instance segmentation of small, densely packed microbial cells, their substantial computational footprint typically makes them too slow. Moreover, their generalist nature might require significant prompting or fine-tuning to achieve the instance-level accuracy needed for reliable single-cell analysis, differing from their typical "segment anything" behavior.
\end{itemize}
 
Given this landscape, and the critical need for solutions that balance high accuracy with very low latency for event-driven workflows in microbial \gls{mlci}, a systematic comparative evaluation of these leading \gls{dl} segmentation approaches—as undertaken in this paper—is essential to identify the most suitable candidates.
This review of the related work highlights the existing gaps and challenges in developing fully integrated, intelligent, and event-driven \glspl{eap} for microbial single-cell analysis, thereby motivating the specific contributions of \gls{eap4emsig} detailed in the subsequent sections.

\section{Methodology}
\label{section:methodology}
\begin{figure*}[tb]
        \centering
        \includegraphics[width=\linewidth]{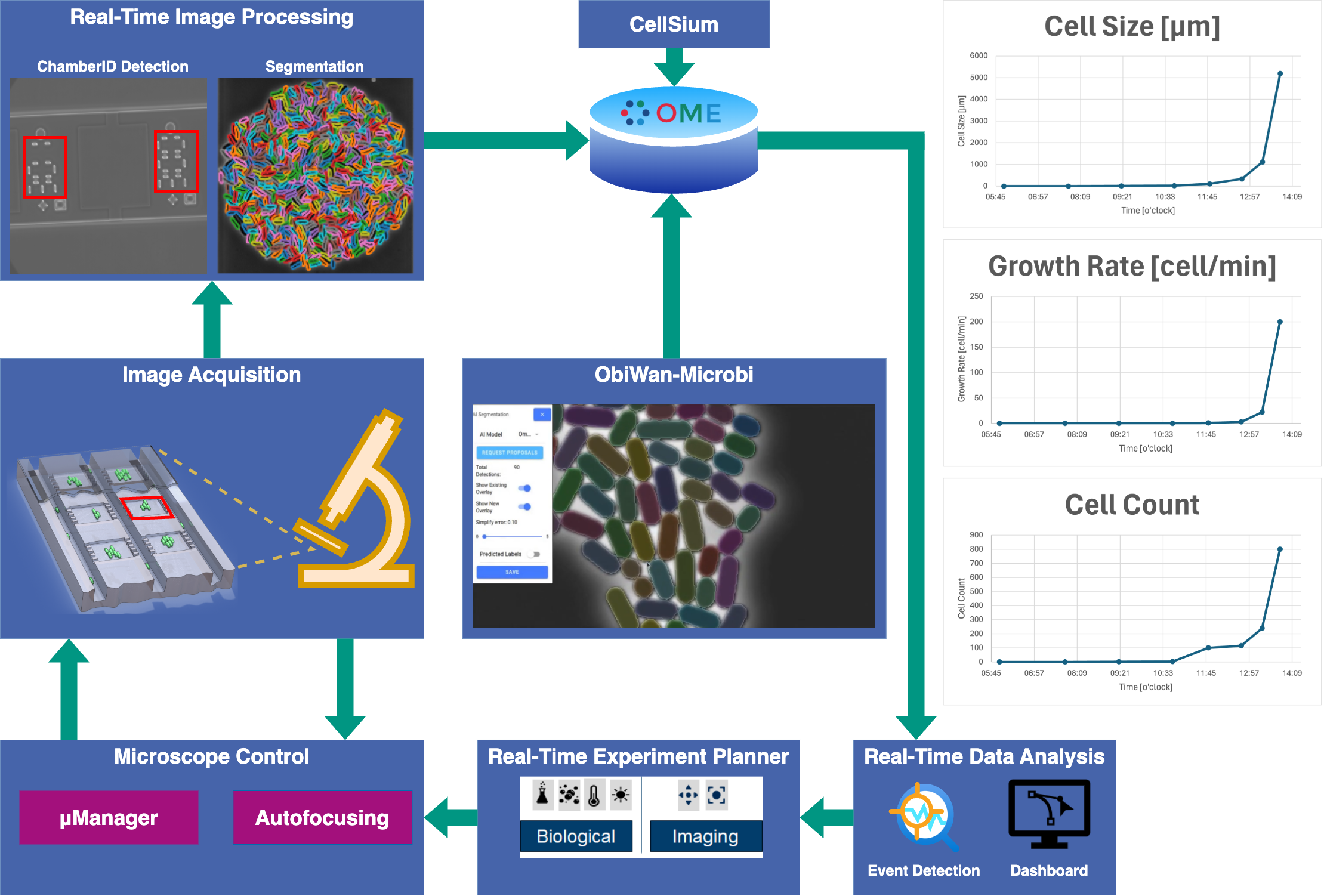}
        \caption{EAP4EMSIG visualization. The pipeline consists of eight modules, represented by the light blue boxes and the \gls{omero} database, arranged in a cyclical process. The microbial images in the figure come from a dataset presented in ~\cite{trackoneinamillion}. The images from the experiment chip are from an internal dataset. Adapted from \cite{friederich2024eap4emsig}.}
        \label{fig:exp_auto_pipeline}
    \end{figure*}

The methods presented in this paper enhance our previously introduced \gls{eap4emsig}~\cite{friederich2024eap4emsig}. This modular pipeline (see \cref{fig:exp_auto_pipeline}), designed for automated, event-driven microfluidic live-cell experiments, comprises eight interconnected modules. Briefly, these include (1) image acquisition; (2) real-time image processing for single-cell instance segmentation; (3) data and metadata management via an \gls{omero} database; (4 \& 5) management of simulated cell data from CellSium~\cite{sachs2022cellsium}  (where ground truth is inherently known) and support for semi-automatic annotation using ObiWan-Microbi~\cite{obiwanMicroby_microbseg}; (6) real-time data analysis with event detection and a dashboard; (7) a real-time experiment planner; and (8) a microscope control module.

Building on our previous evaluation of robotic middleware~\cite{friederich2024eap4emsig}, we have now effectively adopted the \gls{dora} as the backbone of our \gls{eap}. The decision to transition from \gls{ros} to \gls{dora} was motivated on the one hand by the complexity of installing and maintaining \gls{ros}~\cite{fischer2020forgotten}, and on the other hand by \gls{dora}'s superior support for asynchronous, low-latency communication between distributed modules, as well as its composable and scalable architecture.
    
This work introduces novel methodologies and evaluations for three critical components within this \gls{eap4emsig} framework:\gls{mlp}-based autofocusing, real-time image processing (specifically, the evaluation of segmentation methods), and real-time data analysis. These are detailed in the following sections.

\subsection{MLP-based Autofocusing} 
\label{subsection:methodology-autofocusing}
%\begin{figure*}[tb]
%    \centering
%    \includegraphics[width=+\linewidth]{Images/autofocussing-pipeline.png}
%    \caption{Overview of the \gls{mlp}-based autofocusing pipeline. The process begins with feature extraction using techniques %such as Laplacian and Gaussian pyramids, wavelets, and Log-Gabor filters. These features are augmented with random flips and rotations to improve robustness. A multi-layer perceptron is then trained on the extracted features to predict the focus offset.}
%    \label{fig:autofocusing_pipeline}
%\end{figure*}

Maintaining optimal focus is essential for acquiring high-quality image data in time-resolved \gls{mlci}. We propose a regression-based deep learning approach that predicts the focus offset from rapidly computable image features, enabling fast and accurate adjustments without the need for additional, specialized microscope hardware.

\subsubsection{Defining "Perfect Focus" and Ground Truth Acquisition}
\label{subsection:methodology-autofocusing-perfect_focus_gt_acquistion}
The ground truth for our model was established in a two-stage process. First, for each field of view, an experienced microscopist manually adjusted the objective to find the visually optimal focus. This position served as the center-point ($z_0$) for the subsequent automated acquisition. A symmetric z-stack was then acquired around this point. For annotation, the mid-frame of this stack was systematically defined as the 'perfect focus' and all other frames were labeled with their known physical distance from this center.

%The foundation of our autofocusing model is a dataset of z-stacks that have been acquired. For each field of view, a z-stack was captured with a defined step size of \SI{0.1}{\micro\meter} over a range of \SIrange{-5}{5}{\micro\meter}. The "perfect focus" or zero-plane ($z_0$) within each z-stack, serving as the ground truth target frame, was manually identified by an experienced microscopist based on visual assessment of image sharpness and contrast of the microbial cells. For any other image in the z-stack at position $z_i$, the ground truth "focus offset" is then $\text{offset}_{\text{GT}} = z_i - z_0$.

%The optical system used (Nikon Plan Apo $\lambda$ 20x/0.75 NA objective, air immersion) has a theoretical depth of field (DOF) of approximately \SI{1.47}{\micro\meter}(calculated using $\lambda = \SI{550}{\nano\meter}$, $n = 1$). Our method aims to predict an offset that brings the focal plane as close as possible to the operator-defined optimal plane. The reported \gls{mae} of \SI{0.105}{\micro\meter} (see Section 4.1.2) indicates the model's precision in predicting this offset. This \gls{mae} is notably smaller than the \SI{0.1}{\micro\meter} z-stack sampling interval. This reflects the model's ability to effectively interpolate the optimal focus position between these discrete sampled planes. 

\subsubsection{Model Architecture and Feature Extraction}
\label{subsection:methodology-autofocusing-model_arch_fe}

We selected an \gls{mlp} as our \gls{nn} architecture, prioritizing both computational efficiency and interpretability. This architecture offers a balance between predictive performance and ease of use, making it accessible for domain experts such as microbiologists, who may wish to understand, troubleshoot, or retrain the model as new data become available. In contrast to more complex architectures like transformers — which, while powerful, often function as “black boxes” and can be challenging for non-specialists to interpret — our choice of an \gls{mlp} facilitates transparent integration into experimental pipelines and supports broader adoption by the scientific community.

Our \gls{mlp} takes as input features from a single image (the current, potentially out-of-focus view). We extract a set of features known to correlate with image focus:

\begin{itemize}
    \item Image Pyramids: Laplacian and Gaussian pyramids capture multiscale edge and intensity information.
    \item Wavelet Transforms: Bi-orthogonal wavelets (types 1.1 and 1.3) analyze texture and detail at different frequencies and orientations.
    \item Image Characteristics: Information about image orientation and resolution are also included as features. 
\end{itemize}

These extracted features form the input vector for the \gls{mlp}. The \gls{mlp} consists of two hidden layers with \gls{relu} activation functions, AdamW optimizer with a learning rate of $3e-4$ and an output layer that predicts a single continuous value: the focus offset (in \SI{}{\micro\meter}), including its direction (positive or negative), required to reach the optimal focal plane.

\subsubsection{Training and Validation}
\label{subsection:methodology-autofocusing-train_val}
To enhance robustness, data augmentation techniques, including random flips and rotations, were applied during training. The model was trained using \gls{mae} as the loss function, and K-fold cross-validation was employed to ensure robustness. In operation, the system employs an iterative process to enhance robustness. This iterative capability enables the system to correct larger initial deviations from focus and compensate for potential focal drift during long-term experiments.

%The \gls{mlp} was trained on a dataset of approximately 13,000 high-resolution images (2560×2160 pixels). It is important to note that these images were acquired under low-variance conditions, originating from a single, long-running experiment type. While this dataset reflects variations in chamber populations and cell orientations as the microbes grow, the overall experimental setup—including the microfluidic chip design, illumination, and temperature—remained consistent. This specific context is crucial for understanding the model's performance and its intended application range. To enhance robustness and generalization, data augmentation techniques, including random flips and rotations (\SI{\pm 10}{\degree}), were applied during training.

%The dataset was split into a training set (\SI{75}{\percent}), a validation set (\SI{15}{\percent}), and a test set (\SI{10}{\percent}). The validation set was used for hyperparameter optimization and to implement early stopping, preventing overfitting. The model was trained using the Adam optimizer with \gls{mae} as the loss function. K-fold cross-validation was also employed during development to ensure model robustness. The training can be performed within a few hours on standard lab computers, making the model accessible and cost-efficient.

\subsubsection{Real-time Operation}
\label{subsection:methodology-autofocusing-rt_operation}
In operation, the trained \gls{mlp} predicts the focus offset from the features of a single, currently acquired image in under \SI{100}{\milli\second}. This predicted offset is then relayed to the microscope control module. While a single prediction is designed to bring the image close to optimal focus, often within the depth of the field, an iterative process is implemented for enhanced robustness. This iterative capability enables the system to reliably correct larger initial deviations from focus and actively compensate for potential focal drift resulting from environmental factors or mechanical settling over extended experimental durations.

\subsection{Real-Time Image Processing}
\label{subsection:methodology-rt_img_proc}
Accurate instance segmentation of individual microbial cells is crucial for extracting quantitative data and detecting cellular events in real-time. Given the time constraints of event-driven microscopy (target processing time $<\SI{100}{\milli\second}$ for this work), this section outlines the methodology for evaluating existing \gls{sota} \gls{dl} segmentation methods rather than developing a new one.

The image processing workflow first involves identifying the current chamber of interest within the acquired image. This can be achieved using established methods, such as classical template matching, or contemporary \gls{dl}-based object detectors, like \gls{yolo}, selected based on the specific experimental setup's speed and accuracy requirements. Microfluidic structures extraneous to the growth chamber are then computationally removed.

Subsequently, the content of the identified chamber, primarily the microbial cells, is segmented to delineate individual cell instances using a suitable segmentation algorithm. Our goal was to identify models that strike a balance between accuracy and low latency. Given the diverse landscape of segmentation models and their varying trade-offs between accuracy, speed, and generalization capabilities for microbial imagery (as discussed in~\cref{subsection:related_work-segmentation}), a systematic benchmark is essential to identify the most suitable candidates for our real-time pipeline. For this, we conducted a benchmark (detailed in~\cref{subsection:experiments-rt_img_proc_seg}) comparing various \gls{sota} \gls{dl} segmentation models (task-specific, domain-specific and foundation models, as reviewed in~\cref{subsection:related_work-segmentation}) on a representative microbial dataset. The evaluation focuses on their zero-shot performance in terms of \gls{sq} (e.g., \gls{pq}) and inference speed, to identify candidates suitable for integration into our real-time \gls{eap4emsig} pipeline.

\subsection{Real-Time Data Analysis}
\label{subsubsection:methodology-rt_data_analysis}
This module leverages the data generated from real-time image processing (e.g., cell masks, cell size, growth rates) to provide immediate insights and enable expert intervention during ongoing experiments. It comprises two main submodules: event detection and a data analysis dashboard.

\subsubsection{Event Detection}
\label{subsubsection:methodology-rt_data_analysis-event_detection}
The event detection submodule is designed to identify predefined biological or technical occurrences at various levels (cultivation chip, chamber or individual cell). Events are defined based on rules specified by domain experts. These expert-defined rules are typically implemented as logical conditions or thresholds applied to quantitative data extracted from the image processing pipeline (e.g., cell count exceeding a defined limit, growth rate changes surpassing a specific speed, or image quality metrics from the autofocus module falling below a set value for a defined duration). These events can be categorized into two main types:

\begin{itemize}
    \item Technical Events: Issues such as focus loss (characterized by persistent large offsets reported by the autofocus system or a decline in image quality metrics below acceptable thresholds), chamber defects (identified through image analysis that assesses the integrity of the chamber, revealing structural issues that may compromise performance), or fluidic anomalies can occur and potentially disrupt the imaging process and affect sample quality.

    \item Biological Events: Significant changes in microbial behavior, such as rapid alterations in growth rates, cell death exceeding a certain percentage (identified via morphological changes or specific stains if used), or specific morphological transitions.
\end{itemize}
 Upon detection of events, the system can trigger automated responses (via the real-time experiment planner module) or notify the user, for instance, via Slack\footnote{https://slack.com/} messages to a dedicated channel, enabling timely intervention.
\subsubsection{Data Analysis Dashboard}
\label{subsubsection:methodology-rt_data_analysis-dashboard}
The dashboard provides a user-friendly interface for biologists to monitor experiments, visualize data in real-time and manage experimental parameters. Key methodological considerations in its design include:

\begin{itemize}
    \item Modular Architecture: Ensuring new features, visualizations, or control elements can be integrated without overhauling the existing codebase, making it adaptable to diverse experimental needs.
    \item Intuitive User Interface: Designed for ease of use by biologists, offering clear visualizations (e.g., heatmaps of chip status, time-series plots of cellular metrics per chamber) and straightforward controls.
    \item Real-time Feedback: Displaying up-to-date information on experiment status, key cellular metrics (cell count, size, growth rate, focus score), and detected events.
    \item Interactive Control: Allowing users to adjust experimental parameters or trigger specific actions based on observed data, facilitating biologist-in-the-loop operation.
\end{itemize}

\section{Experiments}
\label{section:experiments}
This chapter presents the experimental test of the three key modules introduced in \cref{section:methodology}. For each module, we describe the experimental setup, present the results and provide a discussion of the findings.

\subsection{MLP-based Autofocusing}
\label{subsection:experiments-autofocusing}

\subsubsection{Experimental Setup}
 The images were captured using a Nikon T1 microscope equipped with a Nikon Plan Apo $\lambda$ 20x/0.75 NA objective (air immersion). For each field of view, a z-stack was acquired with a defined step size of \SI{0.1}{\micro\meter} over a range of \SI{-5}{\micro\meter} to \SI{+5}{\micro\meter}. It is essential to note that these images were acquired under low-variance conditions, originating from a single, long-running experiment type where factors such as the microfluidic chip design, illumination, and temperature remained consistent. To ensure the physical capability for fine-grained adjustments, we confirmed with the manufacturer (Nikon) that the microscope's z-drive features a mechanical step resolution of \SI{0.025}{\micro\meter} and a positioning accuracy of \SI{0.065}{\micro\meter}. Initially, 5\% of the total 13,000 high-resolution images (2560×2160 pixels) were reserved for testing the autofocusing \gls{mlp}-based model (see~\cref{subsection:methodology-autofocusing}) using a stack-based splitting strategy to prevent data leakage, ensuring that all frames from the same experiment remained grouped together. Subsequently, from the remaining 95\% of the dataset, another split was applied, resulting in 76\% of the total data used for training and 19\% for validation.

Model performance was primarily assessed using the \gls{mae} by comparing the predicted z-offset values with the ground truth z-offsets. All models were implemented using TensorFlow/Keras and trained and evaluated on an NVIDIA RTX 3090 GPU, reflecting a realistic laboratory hardware deployment.

\subsubsection{Results and Discussion}

\begin{figure}[h]
    \centering
    \includegraphics[width=0.99\linewidth]{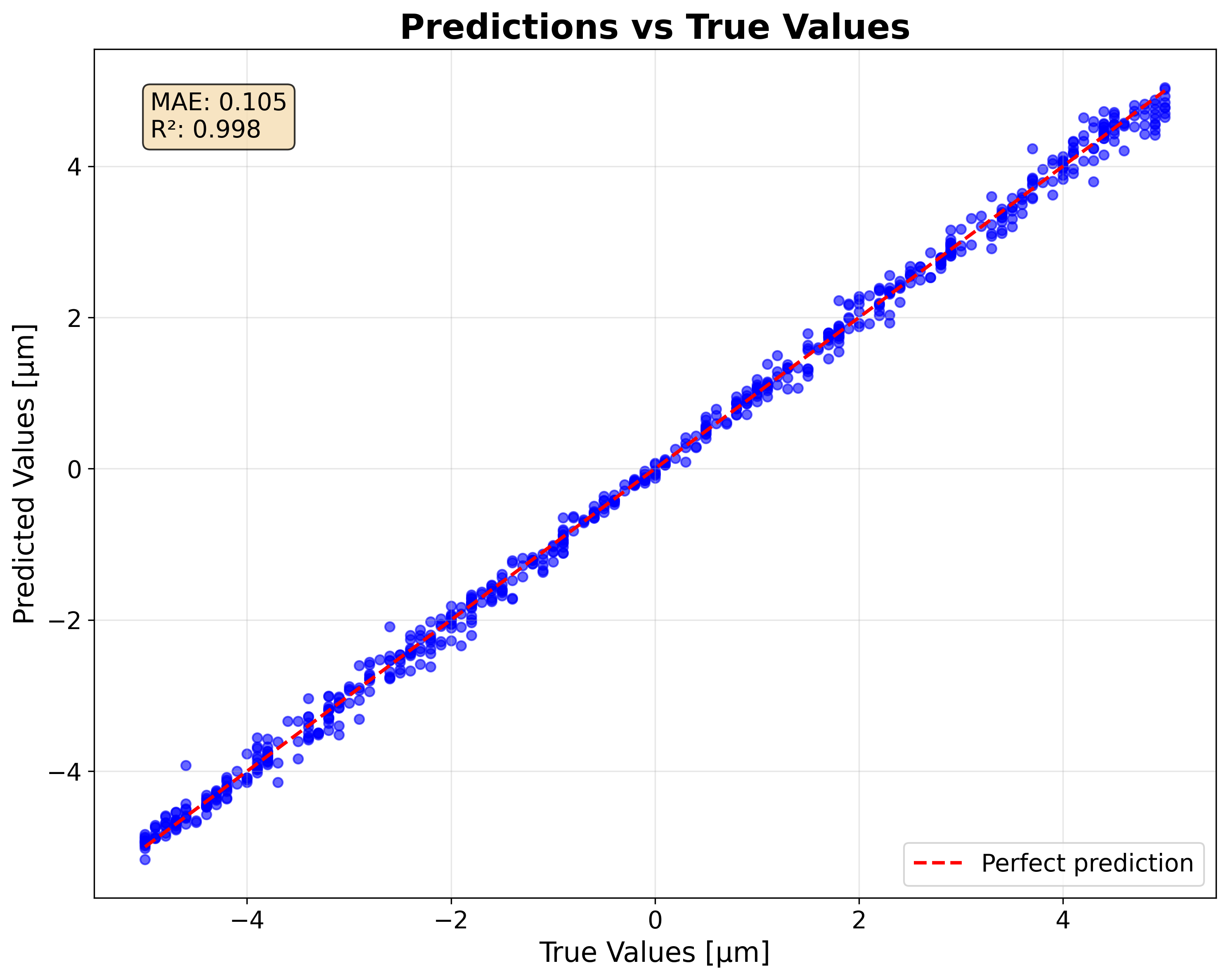}
    \caption{Comparison of predicted and actual focus offset values. The scatter plot shows the focus offset predicted by the \gls{mlp} model (Y-axis) in relation to the ground truth values (X-axis) for the test dataset. The red dashed line represents a perfect match. The model achieves a high determination coefficient $R^2$ of 0.998 and a \gls{mae} of \SI{0.105}{\micro\meter}.}
    \label{fig:experiments-autofocusing_predictions-true-values}
\end{figure}

\begin{figure}[h]
    \centering
    \includegraphics[width=0.99\linewidth]{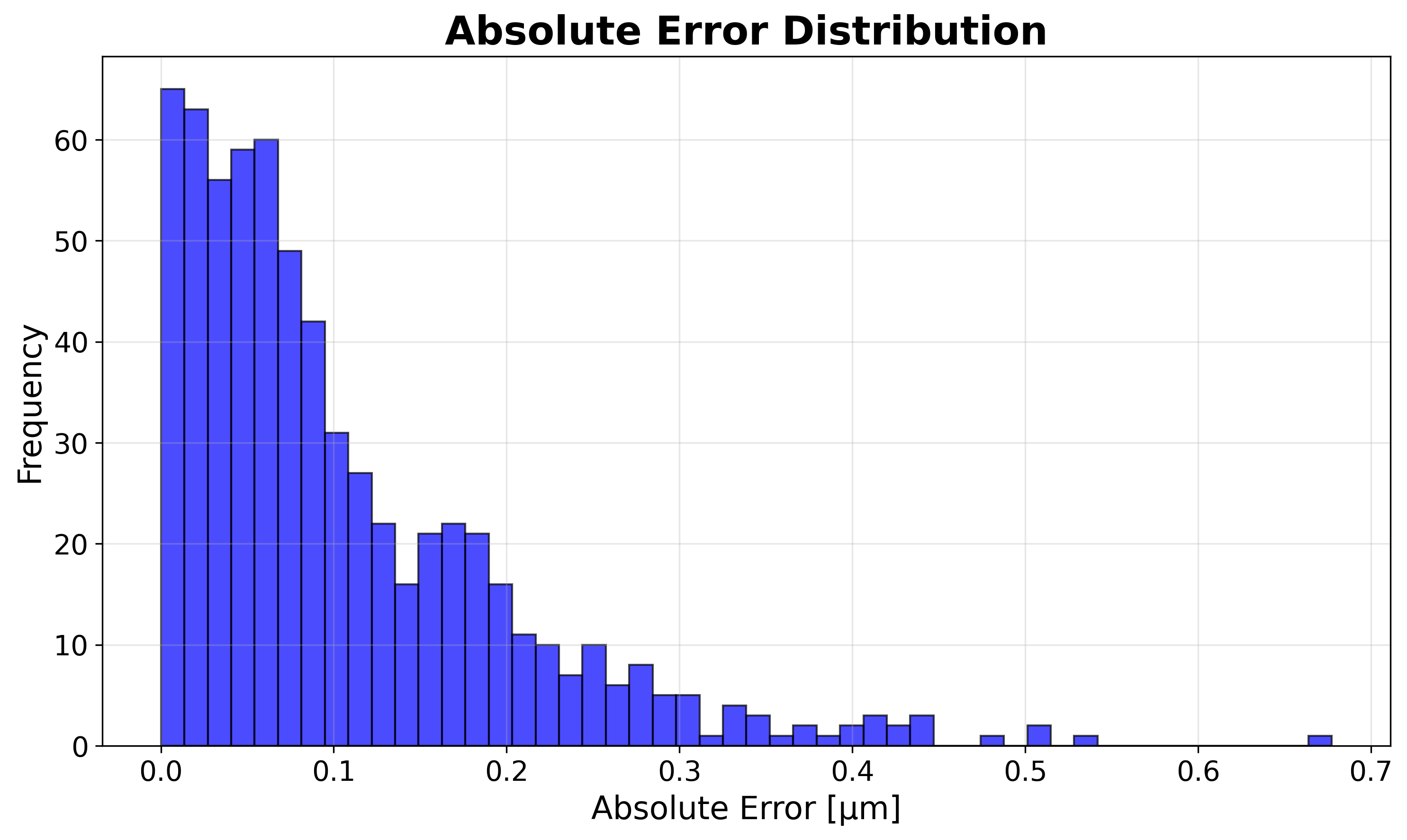}
    \caption{Distribution of the absolute prediction error. The histogram shows the frequency distributions of absolute errors on the test dataset. The right-skewed distribution illustrates that the majority of prediction errors are very small, clustering near zero, while larger errors are infrequent.}
    \label{fig:experiments-autofocusing_error-percentilies}
\end{figure}

\begin{figure}[h]
    \centering
    \includegraphics[width=0.99\linewidth]{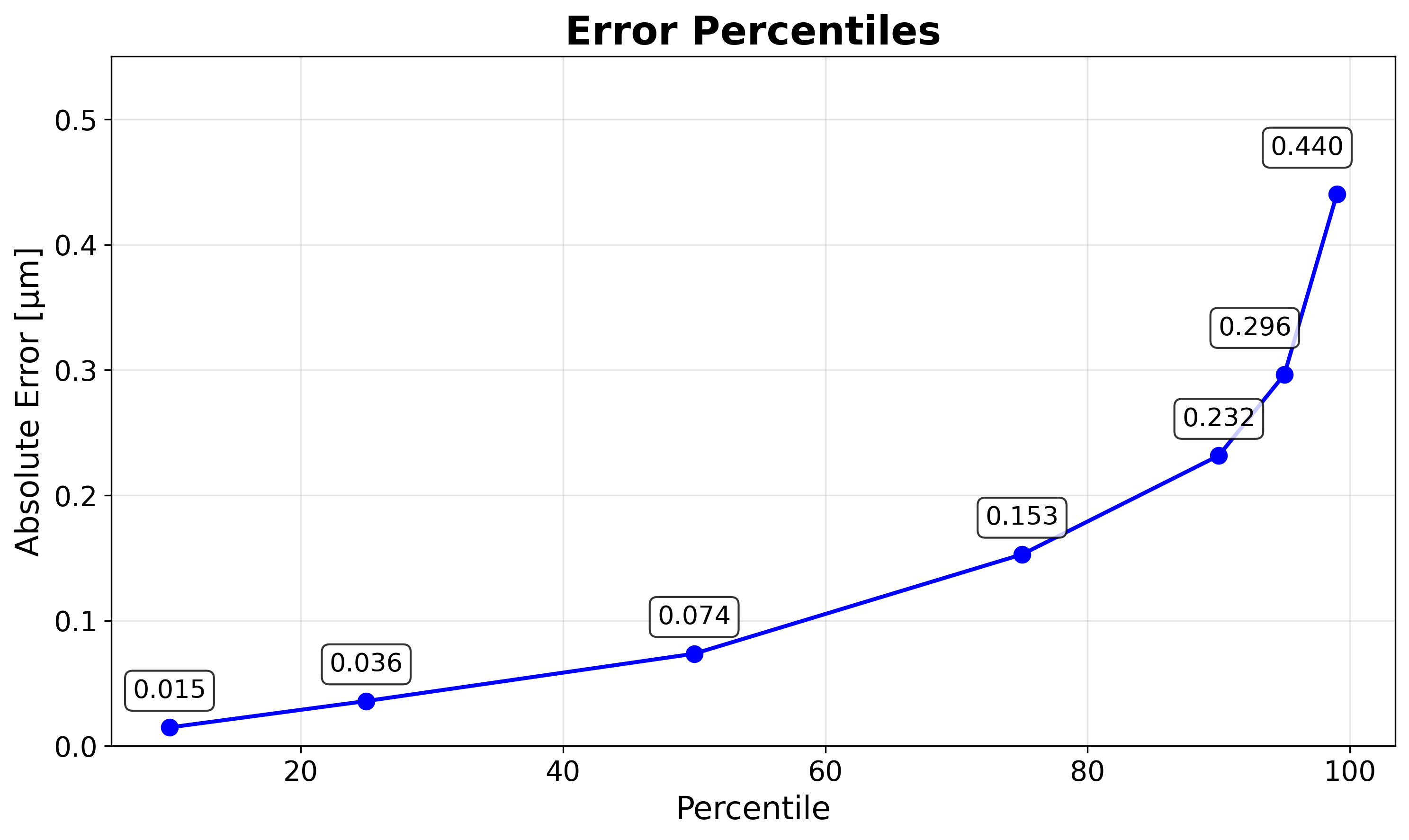}
    \caption{Absolute prediction error by percentile. The figure quantifies the error distribution on the test dataset. The curve shows the absolute error values for each percentile. For instance, the median error (50th percentile) is \SI{0.074}{\micro\meter}, while 75\% of all errors are below \SI{0.153}{\micro\meter}. The 99th percentile is \SI{0.44}{\micro\meter}.}
    \label{fig:experiments-autofocusing_absolute-error-distribution}
\end{figure}

The performance of the trained \gls{mlp}-based autofocusing model on the independent test set is detailed in \cref{fig:experiments-autofocusing_predictions-true-values,fig:experiments-autofocusing_absolute-error-distribution,fig:experiments-autofocusing_error-percentilies}. This scatter plot of predicted values versus ground truth focusing offset (see~\cref{fig:experiments-autofocusing_predictions-true-values}) demonstrates a strong linear correlation. The model achieved a high coefficient of determination $R^2$ of $0.998$, indicating that the predictions align closely with the ideal one-to-one relationship. The overall \gls{mae} for the test set was \SI{0.105}{\micro\meter}. These results indicate a precise system, with the reported \gls{mae} being notably smaller than the \SI{0.1}{\micro\meter} z-stack sampling interval used for training. This sub-step-size accuracy is possible because the MLP, trained as a regressor, learns to interpolate the optimal focus position by analyzing continuous changes in image features. This predictive capability is physically supported by the microscope's hardware, which has a positioning accuracy (\SI{0.065}{\micro\meter}) finer than the sampling interval.

A more detailed analysis of the prediction error is provided in \cref{fig:experiments-autofocusing_absolute-error-distribution,fig:experiments-autofocusing_absolute-error-distribution}. The histogram of absolute errors (see~\cref{fig:experiments-autofocusing_error-percentilies}) shows a right-skewed distribution, with the vast majority of errors being very small, most \SI{0.2}{\micro\meter}, confirming that substantial prediction errors are rare.

The \gls{mlp}-based autofocusing model, trained as described in~\cref{subsection:methodology-autofocusing-train_val}, achieved a final test \gls{mae} of \SI{0.105}{\micro\meter} with a standard deviation of \SI{0.15}{\micro\meter}. The 25th, 50th (median), and 90th percentiles of the absolute error were \SI{0.036}{\micro\meter}, \SI{0.074}{\micro\meter}, and \SI{0.232}{\micro\meter}, respectively, with a 99th percentile of \SI{0.440}{\micro\meter} and a maximum observed error of \SI{0.68}{\micro\meter} (see~\cref{fig:experiments-autofocusing_error-percentilies}). These results indicate a generally precise system: most prediction errors fall well below \SI{0.153}{\micro\meter}, which is tolerably close to z-stack acquisition step size of \SI{0.1}{\micro\meter} and within a fraction of the microscope’s mechanical step resolution of \SI{0.025}{\micro\meter} (see~\cref{fig:experiments-autofocusing_absolute-error-distribution}).

%\begin{figure*}[h]
%    \centering
%    %\begin{subfigure}[b]{0.3\textwidth}
%    %    \includegraphics[width=\linewidth{Images/mae_regressor_highres_flipOn_bs8_normalized.png}
%    % \caption{Average \gls{mae} per test step.}
%     %   \label{fig:autofocusing_training-mae_regressor_highres_flipOn_bs8_normalized}
%    %\end{subfigure}
%    \begin{subfigure}[b]{0.3\textwidth}
%        \includegraphics[trim={5mm 10mm 5mm 5mm},width=\textwidth]{Images/offsets_v3.png}
%        \caption{\gls{mae} Across Ground Truth Focus Offsets.}
%        \label{fig:autofocusing_training-error_across_offsets}
%    \end{subfigure}
%    %\\begin{subfigure}[b]{0.3\textwidth}
%    %\    \includegraphics[width=\linewidth]{Images/error_histogram.png}
%    %\    \caption{Histogram of Test \gls{mae} Across All Focus Offsets ($\mu$m)}
%    %\    \label{fig:autofocusing_training-error_histogram}
%    %\\end{subfigure}
%    \begin{subfigure}[b]{0.3\textwidth}
%        \includegraphics[trim={5mm 10mm 5mm 5mm},width=\textwidth]{Images/negatives_v3.png}
%        \caption{Histogram of Test \gls{mae} for Negative Focus Offsets [$\mu$m]}
%        \label{fig:autofocusing_training-negative_error_histogram}
%    \end{subfigure}
%    \begin{subfigure}[b]{0.3\textwidth}
%        \includegraphics[trim={5mm 10mm 5mm 5mm},width=\textwidth]{Images/positives_v3.png}
%        \caption{Histogram of Test \gls{mae} for Positive Focus Offsets [$\mu$m].}
%        \label{fig:autofocusing_training-positive_error_histogram}
%    \end{subfigure}
%    \caption{Autofocusing test evaluations.}
%    \label{fig:autofocusing_results}
%\end{figure*}

In terms of speed, the model predicts focus adjustments with an average inference time of \SI{87}{\milli\second} per image on the GPU, meeting our real-time requirements. This performance, combined with the low \gls{mae}, underscores the model's suitability for closed-loop focus control in high-throughput \gls{mlci} experiments.

The presented low \gls{mae} and high speed demonstrate the model's effectiveness for its intended application. However, it is crucial to frame these results within the context of the model's design and training data. The \gls{mlp}-based approach was trained on data with low experimental variance (see~\cref{subsection:methodology-autofocusing-train_val}) and therefore performs as an experiment-specific model.

Its primary strength lies in maintaining focus during long-running experiments where the core setup (e.g., chip type, illumination, temperature) remains stable. In such scenarios, the model can reliably compensate for focus drift over hours or days. Conversely, the model is not designed to generalize across different experimental setups. If a user changes the chip or significantly alters other environmental conditions, its performance would likely degrade, necessitating retraining. This highlights a deliberate trade-off: the current \gls{mlp} provides a real-time, computationally inexpensive solution for stable experiments, while a more general-purpose autofocusing tool would require more complex models (e.g., CNNs) and a significantly more diverse training dataset.

\subsection{Real-Time Image Processing: Segmentation}
\label{subsection:experiments-rt_img_proc_seg}

\subsubsection{Experimental Setup}
\label{subsubsubsection:experiments-rt_seg-dataset_metric_impl}
The performance of eleven \gls{sota} \gls{dl} segmentation methods (as categorized in~\cref{subsection:related_work-segmentation}) was evaluated on the benchmark dataset presented by Seiffarth et al.~\cite{trackoneinamillion, dataMicrobsTrackingMillion}, which contains 4000 images of Corynebacterium glutamicum microcolonies from five video sequences, representing typical \gls{mlci} experiments. Ground truth instance segmentation masks are provided with the dataset. The benchmark was performed on an Ubuntu 22.04 workstation with an Intel Core i9-13900 CPU, an NVIDIA RTX 3090 GPU, and 64 GB RAM. All models were evaluated using their default settings to ensure a fair comparison of their out-of-the-box capabilities.

Segmentation accuracy was assessed using \gls{ap}, including \gls{ap}@50 and \gls{ap}@75 and \gls{pq}~\cite{kirillov2019panoptic}, which comprises \gls{sq} and \gls{rq}. These metrics were calculated using TorchMetrics. As \gls{ap}-based metrics require confidence scores, they could not be computed for all evaluated methods. Average inference times were measured using 32-bit floating-point precision, defined as the duration from inputting the image to receiving the model's prediction as an instance mask, including any necessary post-processing.

\subsubsection{Results and Discussion}
\label{subsubsubsection:experiments-rt_seg-results}

\begin{figure*}[h]
    \centering
    \begin{subfigure}[b]{0.16\textwidth}
        \includegraphics[width=\linewidth]{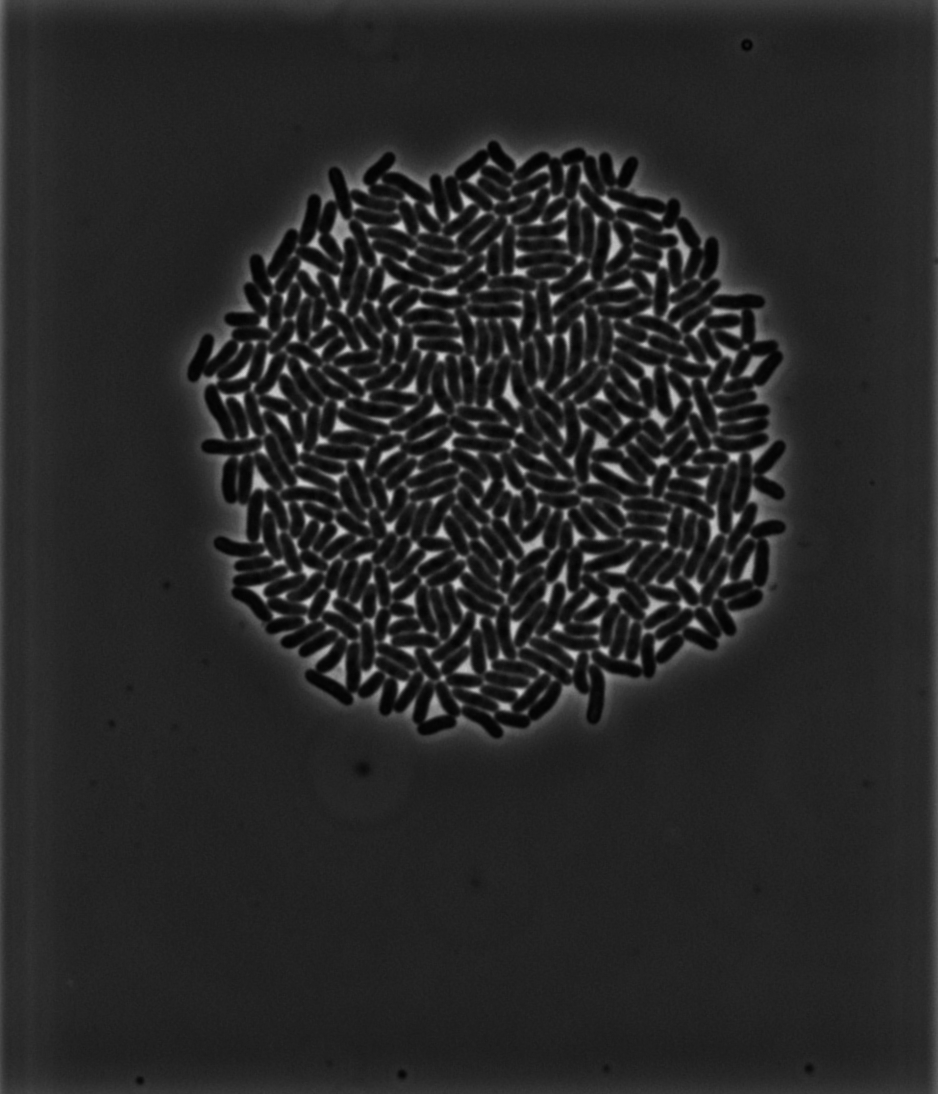}
        \caption{Original}
        \label{fig:results_org}
    \end{subfigure}
    \begin{subfigure}[b]{0.16\textwidth}
        \includegraphics[width=\linewidth]{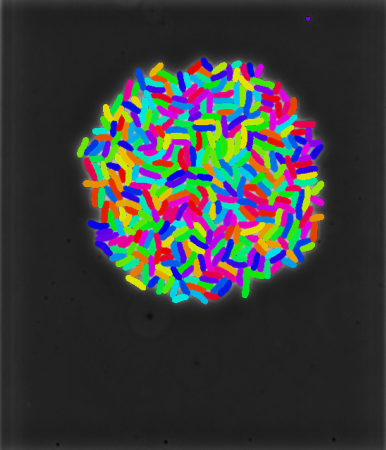}
        \caption{Omnipose~\cite{cutler2022omnipose}}
        \label{fig:results_omnipose}
    \end{subfigure}
    \begin{subfigure}[b]{0.16\textwidth}
        \includegraphics[width=\linewidth]{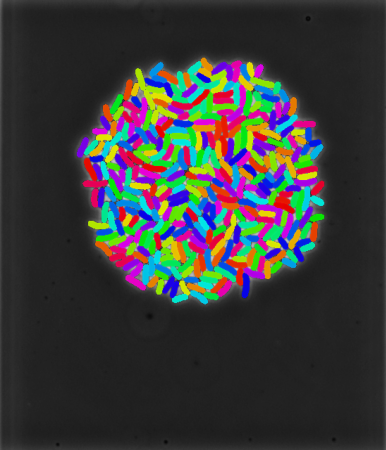}
        \caption{Distance~\cite{scherr2020cell}}
        \label{fig:results_sam}
    \end{subfigure}
    \begin{subfigure}[b]{0.16\textwidth}
        \includegraphics[width=\linewidth]{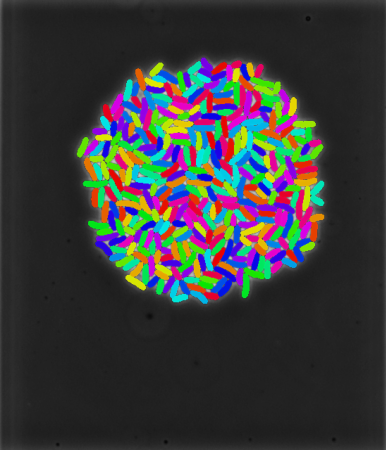}
        \caption{CellPose 3~\cite{Stringer2024.02.10.579780}}
        \label{fig:results_stardist}
    \end{subfigure}
    \begin{subfigure}[b]{0.16\textwidth}
        \includegraphics[width=\linewidth]{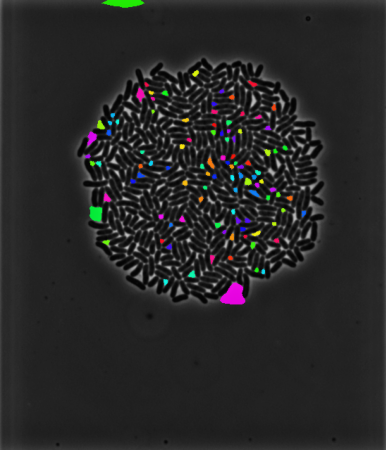}
        \caption{StarDist~\cite{schmidt2018cell}}
        \label{fig:results_stardist}
    \end{subfigure}
    \begin{subfigure}[b]{0.16\textwidth}
        \includegraphics[width=\linewidth]{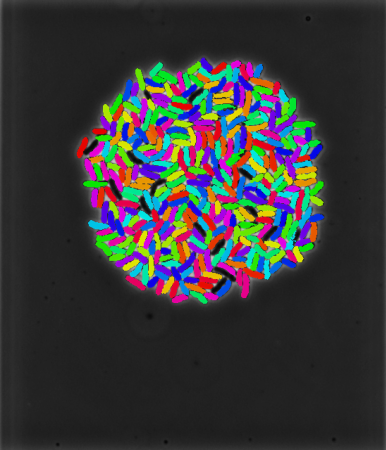}
        \caption{CPN~\cite{celldetection}}
        \label{fig:results_cpn}
    \end{subfigure}
    \begin{subfigure}[b]{0.16\textwidth}
        \includegraphics[width=\linewidth]{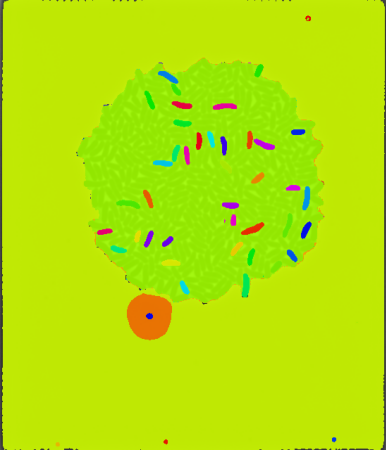}
        \caption{SAM~\cite{kirillov2023segany}}
        \label{fig:results_sam}
    \end{subfigure}
    \begin{subfigure}[b]{0.16\textwidth}
        \includegraphics[width=\linewidth]{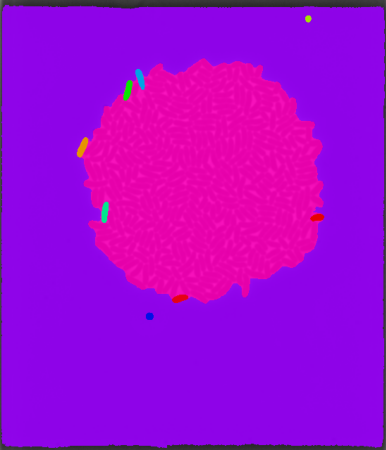}
        \caption{SAM~2~\cite{ravi2024sam}}
        \label{fig:results_sam2}
    \end{subfigure}
    \begin{subfigure}[b]{0.16\textwidth}
        \includegraphics[width=\linewidth]{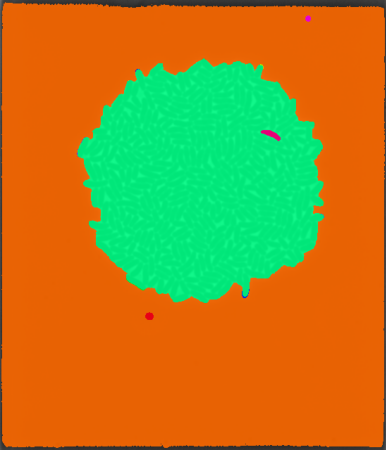}
        \caption{SAM~2.1~\cite{ravi2024sam}}
        \label{fig:results_sam2.1}
    \end{subfigure}
    \begin{subfigure}[b]{0.16\textwidth}
        \includegraphics[width=\linewidth]{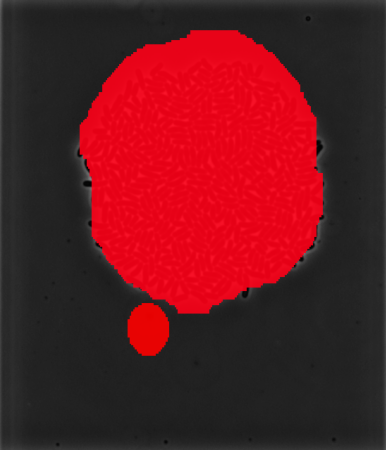}
        \caption{4M21~\cite{4m21}}
        \label{fig:results_4m21}
    \end{subfigure}
    \begin{subfigure}[b]{0.16\textwidth}
        \includegraphics[width=\linewidth]{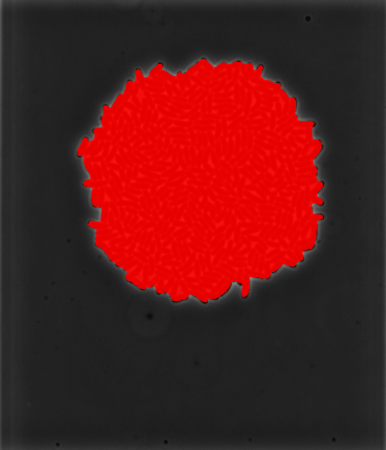}
        \caption{Florence~2~\cite{xiao2024florence}}
        \label{fig:results_florence2}
    \end{subfigure}
    \begin{subfigure}[b]{0.16\textwidth}
        \includegraphics[width=\linewidth]{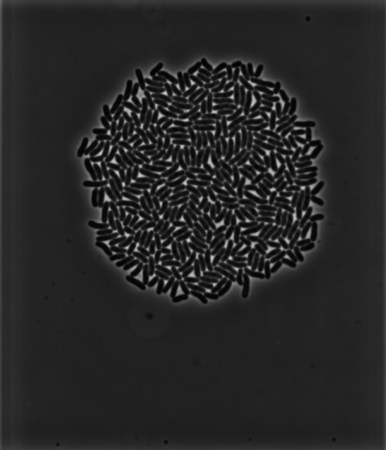}
        \caption{BiomedParse~\cite{zhao2024biomedparse}}
        \label{fig:results_biomedparser}
    \end{subfigure}
    
    \caption{The original image (\cref{fig:results_org}) and the zero-shot instance segmentation predictions for one sample image from~\cite{trackoneinamillion}  (\cref{fig:results_omnipose} to~\cref{fig:results_biomedparser})}.
    \label{fig:results_seg}
\end{figure*}

The qualitative and quantitative results of the segmentation benchmark are presented in~\cref{fig:results_seg}  and~\cref{tab:experiments-od_methods-repro-results}, respectively.

\begin{table*}[tb]
    \centering
    \caption{\gls{ap} results, \gls{pq} results comprising \gls{sq} score and \gls{rq} score as well as inference times (Inf.)  evaluated on the benchmark dataset~\cite{trackoneinamillion} best in bold. When calculating the metric, falsely detected backgrounds were not removed and evaluated as false positives during the \gls{ap} calculation. The models were used according to basic configurations for fair comparison. The values in bold are the best across all methods, provided that results were available. To ensure a fair comparison, we define inference time as the duration from inputting the image to receiving the model's prediction as an instance mask with confidence scores. This includes post-processing needed by certain methods, such as converting predicted contours to a pixel-wise mask. The inference time is measured using 32-bit floating-point precision.}
    \begin{tabular}{l||c|c|c||c|c|c||c}
        \toprule
        Methods \textbackslash Metrics & AP$\uparrow$~[\%] & AP@50$\uparrow$~[\%] & AP@75$\uparrow$~[\%] & PQ$\uparrow$~[\%] & PQ-SQ$\uparrow$~[\%] & PQ-RQ$\uparrow$~[\%] & $\varnothing$Inf.~[ms/img]$\downarrow$\\
        \midrule
        Omnipose~\cite{cutler2022omnipose} & - & - & - & 93.36 & 93.95 & 99.35 & 271\\
        Distance-based~\cite{scherr2020cell} &- &- &- &93.02 &93.48 &\textbf{99.50} & \textbf{121}\\
        \midrule
        Cellpose~3~\cite{Stringer2024.02.10.579780} &-&-&- & \textbf{93.58} & \textbf{94.07} &99.46 &1,115\\
        StarDist~\cite{schmidt2018cell} & 0 & 0 & 0 & 36.29 & 72.87 & 40.93 & 7,686\\
        CPN~\cite{celldetection} & \textbf{62.32} & \textbf{95.51} & \textbf{81.70} & 85.75 & 87.79 & 97.63 & 185\\
        \midrule
        SAM~\cite{kirillov2023segany} & 3.47 & 4.76 & 4.70 &6.26 & 84.16 & 7.36 & 1,994\\
        SAM~2~\cite{ravi2024sam} &0.27 &0.33 &0.32 &4.85 &78.49 &6.89 & 1,566\\
        SAM~2.1~\cite{ravi2024sam} &1.79 &2.64 &2.45 &6.00 &76.10 &7.98 & 1,546\\
        \midrule
        4M21~\cite{4m21} &- &- &- &38.11 &47.80 &39.86 & 103,025\\
        \midrule
        Florence~2~\cite{xiao2024florence} &- &- &- &42.00 &83.78 &43.16 & 4,294\\
        \midrule
        BiomedParse~\cite{zhao2024biomedparse} &0.03 &0.07 &0 &37.63 &78.48 &41.12 & 266\\
        \bottomrule
    \end{tabular}
    \label{tab:experiments-od_methods-repro-results}
    \vspace{-0.2cm}
\end{table*}

Among the evaluated methods, Cellpose 3 achieved the highest \gls{pq} of \SI{93.58}{\percent} and RQ of \SI{99.46}{\percent}, largely due to its automatic cell diameter estimation. However, this came at the cost of significantly longer inference times (\SI{1115}{\milli\second}), making it nearly ten times slower than the Distance-based method. The Distance-based method, while achieving a slightly lower \gls{pq} of \SI{93.02}{\percent}, was the fastest among the highly accurate models at \SI{121}{\milli\second}. This substantial speed advantage is likely due to its comparatively simpler model architecture, which requires fewer computational steps than the more complex network used by Cellpose 3. Omnipose also performed well with a \gls{pq} of \SI{93.36}{\percent} and an inference time of \SI{271}{\milli\second}. Visually, the results from Cellpose 3, Omnipose and the Distance-based method were nearly indistinguishable, all reliably detecting microcolonies and their constituent cells. Given its speed and high accuracy, the Distance-based method presents the best option for real-time applications, followed closely by Omnipose.

As anticipated, the \glspl{fm} (\gls{sam},~\gls{sam}~2,~\gls{sam}~2.1,~4M21,~Florence-2,~BiomedParse) were generally unsuitable for this specific real-time instance segmentation task. They primarily identified the microcolony as a whole rather than resolving individual cells, and their inference times were often unacceptably high (e.g., $ > \SI{100}{\second}$ for 4M21). This is likely due to these models being trained on vastly different and broader datasets, not optimized for the fine-grained instance segmentation of small, densely packed microbial cells without specific prompting or fine-tuning. BiomedParse, trained on medical objects like organs, failed to segment microcolonies effectively. StarDist, optimized for bright objects on dark backgrounds, struggled with the dark microcolonies against a dark background in our dataset. \gls{cpn}, while faster than Cellpose 3, had difficulty with densely packed regions.

Regarding the \gls{pq} metric, we acknowledge that for very small and densely packed objects like bacteria, \gls{pq} can be sensitive to minor contour inaccuracies, potentially impacting the absolute scores~\cite{foucart2023panoptic}. However, \gls{pq} is a widely adopted metric in cell segmentation benchmarks, and its components (\gls{sq} and \gls{rq}) provide valuable insights into both segmentation and detection aspects~\cite{9325955,9446924}. Furthermore, the relative performance rankings and the substantial differences in inference times observed in our benchmark remain informative for selecting suitable models. The \gls{ap} metrics reported in~\cref{tab:experiments-od_methods-repro-results}, where available, offer an additional perspective on performance.

This benchmark highlights the critical trade-off between segmentation quality and inference speed, emphasizing the need to select models based on specific experimental requirements for real-time, event-driven \gls{mlci}.

\begin{figure*}[tb]
        \centering
        \includegraphics[width=\linewidth]{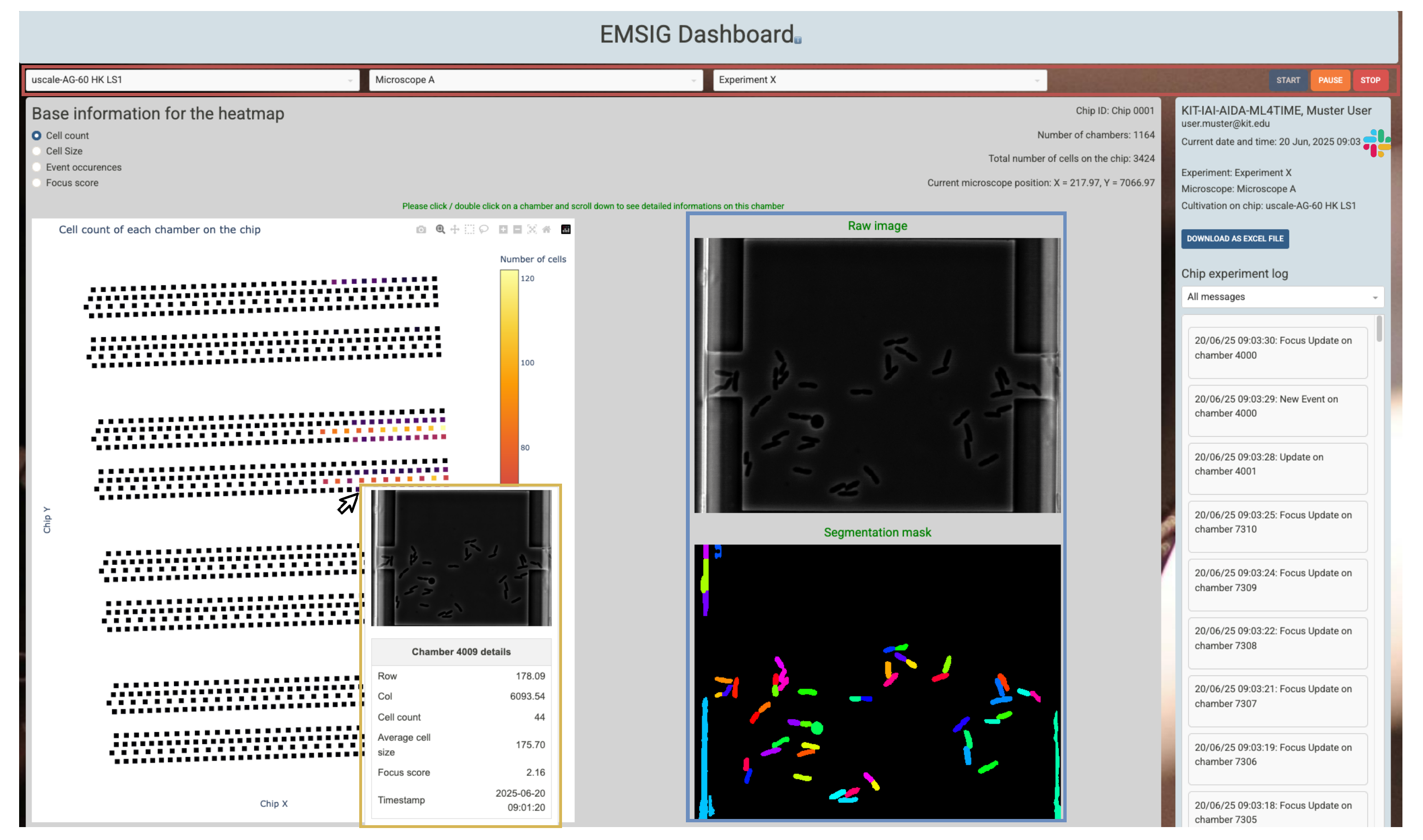}
        \caption{\gls{eap4emsig} Dashboard: Experiment monitoring interface. The primary interface for monitoring experiments, highlighted in the red box, allows users to select the cultivation chip and microscope to initiate an experiment or load a predefined protocol. The heatmap, structured according to the cultivation chip's layout, visually represents the status of each cultivation chamber based on selected metrics, here cell count. The yellow box contains the context window displayed upon hovering on the chamber. Clicking on a chamber opens a detailed view (blue box), which includes the raw region of interest and segmented image.}
        \label{fig:dashboard_heatmap}
\end{figure*}

\begin{figure*}[tb]
        \centering
        \includegraphics[width=\linewidth]{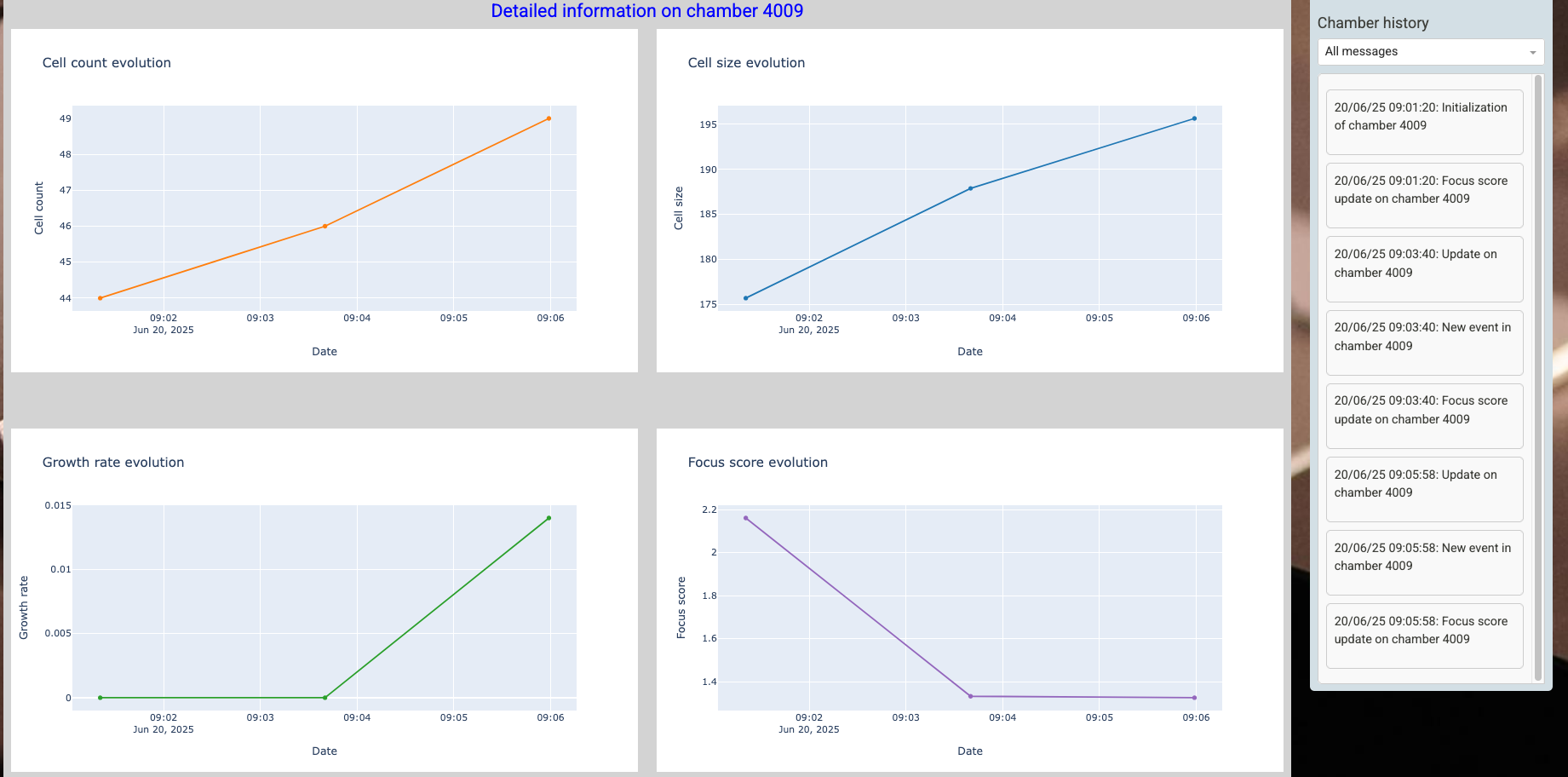}
        \caption{Detailed View of a selected Chamber. The view displays time-series data for key metrics such as cell count, cell size, growth rate, and focus score, providing users with comprehensive insights into the chamber's performance over time.}
        \label{fig:dashboard_chamber_info}
\end{figure*}

\subsection{Real-Time Data Analysis}
\label{subsection:experiments-rt_data_analysis} 
The real-time data analysis module, encompassing event detection and an interactive dashboard (described methodologically in~\cref{subsubsection:methodology-rt_data_analysis}), was tested in experimental runs to demonstrate its functionality. To achieve this, a time-lapse experiment was conducted to grow E. coli microcolonies, capturing a series of images over time. The acquired images were then replayed through the image acquisition module at a constant frame rate, effectively simulating live experiments. This replay served as the entry point for the \gls{eap4emsig} pipeline, enabling us to systematically test and validate each component of the pipeline under conditions that closely mimic real-time operation. Pre-defined triggers for common biological and technical events related to the performed experiment were established.

\subsubsection{Event Detection Demonstration}
\label{subsection:experiments-rt_data_analysis_event_detection} 
The event detection submodule (\cref{subsubsection:methodology-rt_data_analysis-event_detection}) was configured with rules to identify exemplary events. For instance:

An out-of-focus event was triggered if the \gls{mlp}-based autofocus module reported a predicted offset greater than \SI{0.5}{\micro\meter} for three consecutive time points.
A "Rapid Growth" event was triggered if the cell count in a specific chamber (derived from the real-time segmentation module) increased by more than \SI{10}{\percent} within consecutive time points. Upon triggering, the system effectively generated and dispatched notifications to our Slack communication channel, showcasing its ability to alert users to critical events in real-time and facilitating prompt intervention.

\subsubsection{Dashboard Functionality Demonstration}
\label{subsubsection:experiments-rt_data_analysis_dashboard}
The \gls{eap4emsig} dashboard (methodology detailed in~\cref{subsubsection:methodology-rt_data_analysis-dashboard}) provides comprehensive visualization and control features, as demonstrated in~\cref{fig:dashboard_heatmap} and~\cref{fig:dashboard_chamber_info}. \cref{fig:dashboard_heatmap} (red box) presents the primary experiment monitoring interface, where users can select the cultivation chip and microscope to either initiate an experiment or load a predefined protocol. Experiment management is facilitated through control buttons labeled "Start", "Pause/Resume", and "Stop". The heatmap, structured according to the cultivation chip's layout, visually represents the status of each cultivation chamber based on selected metrics, including cell count, cell size, event occurrences, and focus score.

Hovering over a chamber on the heatmap reveals a context window that displays the current visual state of the chamber along with additional details (\cref{fig:dashboard_heatmap} yellow box). As illustrated in~\cref{fig:dashboard_heatmap} and~\cref{fig:dashboard_chamber_info}, clicking on a chamber opens a detailed view. This view includes the raw region of interest from the image alongside the segmented image highlighting individual cells (\cref{fig:dashboard_heatmap}, blue box). Additionally, time-series data for key metrics, including cell count, cell size, growth rate, and focus score, are presented. These plots visualize the expected dynamics of microbial growth, such as the exponential increase in cell count, while derived metrics like the growth rate naturally exhibit higher volatility. A sidebar displays experiment metadata and allows for data export. This empowers users to track the history of individual chambers and filter messages, enabling in-depth analysis of ongoing experiments and supporting informed decision-making.

The presented functionalities demonstrate the dashboard's capacity to provide biologist-in-the-loop control and real-time insights, significantly enhancing the ability to oversee and interpret complex \gls{mlci} experiments efficiently. The pipeline has been in routine operation since 01/2025 and is being continuously expanded based on user feedback.

\section{Conclusion}
\label{section:conclusion}
This paper presents an in-depth exploration and enhancement of three key modules within the \gls{eap4emsig} system, aiming to advance real-time, event-driven microscopy for microfluidic single-cell analysis.

First, we introduced a novel real-time autofocusing module based on a computationally efficient \gls{mlp}. This module achieves a \gls{mae} of~\SI{0.105}{\micro\meter} with prediction times under \SI{87}{\milli\second} on standard CPU hardware. These results underscore its capability for precise and rapid image acquisition, which is critical for effective downstream image analysis in high-throughput experiments. We have also discussed the importance of understanding this \gls{mae} in the context of the optical system's depth of field and the methodology for defining ground truth. Future research will focus on developing models that are even more robust to variations in experimental conditions, potentially requiring minimal retraining, and further optimizing inference speed for live imaging applications. The modular design ensures that as improved autofocus models become available, they can be integrated with minimal changes to the overall framework.

Second, we conducted a comprehensive zero-shot benchmark of eleven \gls{sota} \gls{dl} segmentation models, encompassing task-specific, domain-specific, and foundation model categories. Our findings reveal that while foundation models demonstrate broad applicability in many domains, they were largely unsuitable for our specific task of real-time, instance-level segmentation of microbial cells, primarily due to low accuracy in resolving individual cells and unacceptably high processing times. In contrast, task-specific models like Omnipose and particularly the Distance-based method, along with the domain-specific model Cellpose 3, delivered excellent segmentation quality with \gls{pq} scores exceeding \SI{93}{\percent}. The Distance-based method distinguished itself with a remarkably fast processing time of \SI{121}{\milli\second}. While we acknowledge the discussions regarding the sensitivity of \gls{pq} for small, dense objects, the relative performance differences and inference speeds observed provide crucial guidance for model selection. Future work will explore methods to automatically select the suitable model based on the recorded data such as in~\cite{fingerprint_marcel, figerprint_lena_maier_heine, yamachui_sitcheu_mlops_2023, godau2024knowledgesilostaskfingerprinting}, strategies to accelerate processing times further, such as model conversion to specialized inference formats (e.g., TensorRT) or quantization to lower-precision formats, aiming to consistently meet the sub-\SI{100}{\milli\second} target for the suitable segmentation model.

Finally, the real-time data analysis module, featuring event detection capabilities and an interactive dashboard, was presented. By providing a user-friendly interface for monitoring, real-time data visualization, and experimental control, this module significantly enhances biologists' ability to oversee and interpret ongoing experiments efficiently. The demonstration highlighted its potential for facilitating biologist-in-the-loop decision-making and reacting to critical events. Future enhancements will include integration with electronic lab notebooks (e.g., eLabFTW) to streamline experiment documentation and reporting.

In summary, the advancements presented for these three modules make a significant contribution to the overarching goal of creating a more intelligent, responsive, and automated pipeline for microfluidic single-cell analysis. Continued research will focus on improving the robustness and accuracy of the autofocusing module, refining strategies for the automated selection of optimal segmentation methods based on image characteristics, further enhancing segmentation speed, and expanding the functionalities of the real-time data analysis platform. New models and pipeline integrations are currently under development and will be detailed in forthcoming publications.

\section{Contributions}
The authors have accepted responsibility for the entire content of this manuscript and approved its submission. We describe here the individual contributions: Conceptualization: NF, AJYS, JS, RM; Methodology: NF, AJYS, EY, DK, HS, JS, KN, RM; Software: NF, AJYS, AN, EY, MB, LS, BA, JS; Investigation: NF, AJYS, JS; Resources: JS, KN; Writing – Original Draft: NF, AJYS, MP, MB, ON, EY, JS; Writing – Review \& Editing: NF, AJYS, AN, MP, EY, MB, LS, BA, TL, ON, NK, HS, JS, KN, RM; Supervision: DK, HS, KN, RM; Project administration: DK, HS, KN, RM; Funding Acquisition: DK, HS, EY, KN, RM.

\section{Acknowledgments}
This work was supported by the President's Initiative and Networking Funds of the Helmholtz Association of German Research Centers [Grant EMSIG ZT-I-PF-04-44]. The Helmholtz Association funds this project under the "Helmholtz Imaging Platform", the authors N.~Friederich, A.~J.~Yamachui~Sitcheu and R.~Mikut under the program "Natural, Artificial and Cognitive Information Processing (NACIP)", the authors N.~Friederich and A.~J.~Yamachui~Sitcheu through the graduate school "Helmholtz Information \& Data Science School for Health (HIDSS4Health)" and the author Johannes Seiffarth through the graduate school "Helmholtz School for Data Science in Life, Earth and Energy (HDS-LEE)".

\bibliographystyle{unsrt}  


\end{document}